\documentclass[a4paper, 11pt]{article}

\usepackage{url}
\usepackage{graphicx,color}
\usepackage[arrow,matrix]{xy}
\usepackage[american]{babel}
\usepackage{listings}
\usepackage{paralist}
\usepackage{amsmath}
\usepackage{tikz}
\usepackage{pgfplots}
\usepackage{xspace}
\usepackage{float}
\usepackage{breakurl}
\usepackage{fullpage}
\usepackage{booktabs}
\usepackage[pdftex, pdfauthor={Julian Backes, Michael Backes, Markus Dürmuth, Sebastian Gerling, Stefan Lorenz}, pdftitle={X-Pire! - A digital expiration date for images in social networks}, bookmarks=true]{hyperref}

\title{\textsf{\textbf{\XPIRE~-- A digital
expiration date for images in social networks}}}

\author{Julian Backes$^{1}$, Michael Backes$^{1,2}$, Markus D\"urmuth$^{3}$\\
Sebastian Gerling$^{1}$, Stefan Lorenz$^{1}$}

\date{}

\newcommand\PUBLISHER{X-pire!-Publisher\xspace}
\newcommand\VIEWER{X-pire!-Viewer\xspace}
\newcommand\KEYSERVER{X-pire!-Keyserver\xspace}
\newcommand\XPIRE{X-pire!\xspace}

\newcommand\PUBLISHERA{publisher\xspace}
\newcommand\VIEWERA{viewer\xspace}
\newcommand\KEYSERVERA{keyserver\xspace}
\newcommand\CAPTCHA{Captcha\xspace}
\newcommand\CAPTCHAS{Captchas\xspace}
\newcommand\EXPDATE{expiration date\xspace}

\definecolor{deepred}{RGB}{215, 25, 28}
\definecolor{lightorange}{RGB}{253, 174, 97}
\definecolor{deepblue}{RGB}{44, 123, 182}
\definecolor{deeppurple}{RGB}{94, 60, 153}

\begin{document}
\setlength{\columnsep}{0.8cm}

\maketitle

\vspace{-1cm}

\begin{center}
  \textit{$^{1}$ Saarland University\\
  $^{2}$ Max Planck Institute for Software Systems\\
   $^{3}$ Ruhr-Universit{\"a}t Bochum}\\
\vspace{0.3cm}
\url{http://www.infsec.cs.uni-saarland.de/projects/forgetful-internet/}
\end{center}

\begin{abstract} The Internet and its current information culture of
  preserving all kinds of data cause severe problems with privacy.
  Most of today's Internet users, especially teenagers, publish
  various kinds of sensitive information, yet without recognizing that
  revealing this information might be detrimental to their future life
  and career. Unflattering images that can be openly accessed now and
  in the future, e.g., by potential employers, constitute a
  particularly important such privacy concern.  We have developed a
  novel, fast, and scalable system called \XPIRE  that
  allows users to set an expiration date for images in social networks
  (e.g., Facebook and Flickr) and on static websites, without
  requiring any form of additional interaction with these web pages.
  Once the expiration date is reached, the images become unavailable.
  Moreover, the publishing user can dynamically prolong or shorten the
  expiration dates of his images later, and even enforce instantaneous
  expiration.  Rendering the approach possible for social networks
  crucially required us to develop a novel technique for embedding
  encrypted information within JPEG files in a way that survives JPEG
  compression, even for highly optimized implementations of JPEG
  post-processing with their various idiosyncrasies as commonly used
  in such networks. We have implemented our system and conducted
  performance measurements to demonstrate its robustness and efficiency. 
\end{abstract}

\section{Introduction}
The past decade has brought dramatic changes in the way we live and
work. The wide acceptance of social networks' free dissemination of
personal information, the proliferation of networked devices that
provide a means for such communication anytime, anywhere, and the
resulting abundance of published information present significant
opportunities.  However, this information culture of acquiring and
preserving all kinds of data also causes severe problems with
privacy. Most of today's Internet users, especially teenagers, publish
various kinds of sensitive information, yet without recognizing that
revealing this information might be detrimental to their future life
and career. Unflattering images -- typically published within social
networks -- that can be openly accessed now and in the future, e.g.,
by potential employers, constitute a particularly important such
privacy concern. This scenario is especially problematic since
published information is cached by search engines, duplicated by
mirrors and aggregators, and thus available essentially forever.

In contrast to the current situation, archiving large amounts of data
was traditionally expensive and cumbersome, and a large-scale
acquisition of sensitive information was close to impossible. In fact,
it was the exception rather than the rule that personal information
was explicitly archived; hence most such information became
unavailable over time. (Moreover, even if photographs were properly
stored, papers were properly filed, etc. these data were typically not
accessible without explicit consent of the user.) The Internet with
its current information culture and its infinite memory thus clashes
with people's established expectation about the life-time of the
information that they divulge to the public. This observation is
substantiated by a work of
Mayer-Sch\"onberger~\cite{schoenberger-07-useful-void}, who nicely
compares modern electronic storage in the age of information
technology with the traditional way in which paper documents are
archived. His major observation was that retrieving information was
traditionally often close to impossible after a certain period of
time, and that this situation fuels the expectations of average users
as far as expiration of data is concerned. In our current information
culture, however, acquiring and preserving large amounts of data is
quick and easy.

The challenge is now to imitate the traditional expiration of data by
developing a digital expiration date that lives up to the demands of
the modern information culture -- both in terms of user privacy and
the seamless integration into common user activities in the Internet,
such as publishing and consuming digital content. If published data
becomes unavailable after a certain time, then this closely resembles
the behavior of the paper-based world and thus brings the situation
in-line with the prevailing user expectations.

\subsection{Our Contribution}
We have developed a novel, fast, and scalable system called \XPIRE
that allows users to set an expiration date for images in social
networks (e.g., Facebook and Flickr) and on static websites, without
requiring any form of additional interaction with these web pages.
Once the expiration date is reached, the images become unavailable.
Moreover, the publishing user can dynamically prolong or shorten the
expiration dates of his images later, and even enforce instantaneous
expiration. In the following, we describe our overall approach and
highlight major design decisions.

\medskip\noindent\textbf{High-level view on the protocol.}
\hspace{0.5mm}We implement the digital expiration date for images as
follows: Images are encrypted by a symmetric key that is stored on a
centralized \KEYSERVERA. Access to the key is granted only until the
\EXPDATE set by the \PUBLISHERA during the encryption process.  After
encrypting an image, it is stored on a webserver to be viewed by the
public. In order to view protected images, the key is retrieved from
the \KEYSERVERA and used to replace the encrypted image by its
decrypted version. The \KEYSERVERA additionally supports management
functionality for already created keys; in particular this allows for
prolonging and shortening the \EXPDATE of keys. The latter can even be
used to let images expire instantaneously.

\medskip\noindent\textbf{The major technical challenge:
  Dealing with JPEGs.}  \hspace{0.5mm}Images are typically stored as
JPEG files in static websites and in social networks. We thus have to
embed encrypted images into JPEG files in a way that adheres to the
JPEG format.  Rendering this approach possible for social networks --
where by far the largest number of sensitive images are published --
is conceptually challenging, since social networks typically re-encode
images using JPEG compressions.  Thus, simply uploading the encrypted
data $c$ would result in the publication of a re-encoded version $c'$,
in which the encryption would be fully destroyed. As a consequence,
nobody would be able to decrypt the image anymore; it would expire
instantaneously. To remedy this situation, we have developed a novel
technique for embedding encrypted information within JPEG files in a
way that survives JPEG compression.  Solving this problem was not only
challenging in theory. The implementations of JPEG used in the
Internet are often highly optimized versions that do not strictly
follow the JPEG standard and thus disregard accuracy in favor of
performance (examples include rounding errors, lossy conversions,
etc.). Our technique needs no explicit support from existing
webservers or social networks, and thus allows for a seamless
integration into the existing infrastructure.

\medskip\noindent\textbf{Mitigating the data duplication
  problem.}  \hspace{0.5mm}When protected images are published using
our approach they can be viewed by the public until the \EXPDATE is
reached. Thus, an attacker will also be able to view these images in
this time period (under the assumption that no additional protection
mechanisms are in place, e.g., images are only visible to the friends
of the user within the privacy settings of Facebook). The attacker is
thus in principle able to store the keys needed for their decryption,
and consequently to use them to decrypt these images even after their
\EXPDATE.  Although this limitation is inherent to the problem, and
thus equally applies to other approaches that strive for a digital
expiration date (see the Related Work section below), we are the first
to consider this case to the best of our knowledge. Using our
centralized approach for the \KEYSERVERA allows us to set up effective
countermeasures to prevent an attacker from acquiring keys in a
large-scale, automated manner. First, in order for users to retrieve a
key, our approach requires them to send a hash of the encrypted image
that they intend to view.  Since computing this hash requires to
download the encrypted image that is much larger than a key, it serves
as convincing evidence that the corresponding band-width was consumed,
thereby rendering a simply key-requesting attack much more costly.
Second, we rate limit the retrieval of a large number of keys.
Finally, our approach enables the \PUBLISHERA to additionally require
users to solve \CAPTCHAS before viewing the image (e.g. one for every
photo album they are viewing). We stress that our protocol, and in
particular its capability to deal with JPEGs, would work as well in a
decentralized setting, see Section~\ref{sec:related-work} on Related
Work.

\medskip\noindent\textbf{Efficient, scalable, one-click
  implementation.}  \hspace{0.5mm}We have implemented our system and
conducted performance measurements to demonstrate its efficiency. 
The software integrates seamlessly into the user's workflow when
surfing on the Internet. The natural solution was to implement the
client applications as browser plug-ins such that the existing
workflow is not interrupted by starting an external application. In
the viewing process no user interaction with the software is required;
to publish an image, the user needs only to drag-and-drop this image
into the application, and enter a desired expiration date.

\subsection{Related Work}
\label{sec:related-work}
Meyer-Sch\"onberger~\cite{schoenberger-07-useful-void} presents a nice
introduction to the problems caused by the infinite memory of our
information culture. He proposes to tag sensitive data with an
\EXPDATE and to require all servers handling such data to obey the
\EXPDATE.  However, many servers would presumably not be interested in
cooperating since their business model relies crucially on the
collection and openly distribution of user data. Also a legal
requirement forcing servers world-wide to obey to delete data seems
out of reach in the near future. Moreover, it is technically difficult
to audit whether the server actually deleted the data. Within the last
years several attempts have ought to solve the problems that arose
with the Internet's infinite memory.

The first group of solutions adds an \EXPDATE to data. It is
implemented by encrypting the data itself with a symmetric encryption
key and restricting the access to these keys afterwards.  We stress
that none of these works is capable of dealing with scenarios where
published data is subsequently manipulated, e.g., re-encoding of JPEGs
as commonly done in social networks such as Facebook. Similarly, the
threat of storing keys during validity has not been considered thus
far.

The first works that pursued this solution path aimed at securely
deleting data including copies in archives. However, these works
mainly target corporate use, and come with different requirements and
design principles that are incompatible with our setting.
In particular, all servers are aware of the encrypted nature of the
data, post-processing of the data is not supported, an adversary
grabbing all keys while the data is available is not considered, and
in some proposals the \KEYSERVERA additionally aids in decrypting the
data. 
The first such system we are aware of
is~\cite{Boneh-1996-revocable-backup}, which provided the basic
principles.  
A prominent system is the
\emph{Ephemerizer}~\cite{Perl-05-ephemerizer,perlman-2005-file-system-assured-delete},
which was later improved~\cite{Nair-2007-pki-ephemerizer}.

\emph{Vanish}~\cite{geambasu-09-vanish} constitutes a recent, promising
approach along similar lines by storing shares of the keys in dynamic
hash tables (DHTs), a data structure that underlies P2P-networks.  The
DHT will by design stop replicating the key after a certain time, so
the key is basically lost after that and the encrypted data becomes
unavailable.
An attack against the proposed implementation was recently
published~\cite{wolchok-2010-unvanish}, using a Sybil attack on the
DHT. An improved design of DHTs should fix this problem, at the cost
of relying on a (slightly) more specialized design.
We stress again that images and post-processing in general is not
supported by this approach. Another difference is that we decided to
ground our approach on a centralized keyserver solution, in order to
mitigate the threat of key storage / data duplication, see
Section~\ref{sec:on_the_centralized_approach}. However, we emphasize
that our proposed techniques for dealing with JPEG post-processing are
applicable to their decentralized approach as well.
Another current limitation of this approach is that the time after
which data is not longer replicated by common implementations is often too
short to be useful (8 hours for the proposed implementation). In case
longer timeouts are needed, keys have to be kept ''alive'' by actively
taking measures.  
  
The \emph{EphCom} system~\cite{ephemeral-data-project-2010} is very
similar to Vanish, but uses a clever trick to store the keys in the
cache of DNS servers, based on the presence of generated hostnames.
Similar to Vanish, post-processing of protecting data and the threat
of retrieving keys during validity is not considered, and for
publishing data that expires at a specific time one needs to find a
(large) number of domains that have the same TTL; their study shows
that TTLs of more than 7 days are rather uncommon.

The second group of solutions aims at securing privacy-sensitive
content published in social networks, but based on a different
assumption:
The central difference is that these approaches store all data on an
external, trusted server, which we believe does not scale reasonably
well given the vast amount of images published every day.
One example is FaceCloak~\cite{luo-2009-facecloak},  
similar approaches include~\cite{lucas-2008-flybynight}
and~\cite{guha-2008-noyb}.

Our techniques for robustly embedding information within JPEG files
furthermore resemble steganographic techniques to embed data into
images and other data formats,
e.g.,~\cite{burnett-2010-chipping-censorship}. Since these
steganographic techniques additionally have to ensure that they embed
information in an undetectable manner, existing solutions do not
achieve sufficiently good data rates to robustly embed encryptions of
high quality images into images that are accepted by social networks.
We thus had to developed our own routine for embedding arbitrary data
into JPEG images with a data rate that is sufficient for social
networks.

\subsection{Paper Outline}
Section~\ref{sec:xpire_overview} provides a general overview of our
approach. The full technical details of the system as well as
necessary background information are provided in
Section~\ref{sec:tech_details} and Section~\ref{sec:jpeg_embedding}.
Our implementation is described in Section~\ref{sec:implementation},
which is followed by the results of an experimental analysis of the
implementation in Section~\ref{sec:exp_analysis}.  We provide a discussion 
on X-pire! and its functionality in Section~\ref{sec:discussion}, before we conclude and
present future work in Section~\ref{sec:concl_future}.

\section{Schematic Overview of \XPIRE}
\label{sec:xpire_overview}
We have developed a novel, fast, and scalable system 
that allows users to set an expiration date for images in social
networks (e.g., Facebook and Flickr) and on static websites, without
requiring any form of additional interaction with these web pages. The
\EXPDATE is chosen by the user who publishes the data. Alternatively, users can
decide to specify an undefined expiration date at the time of
publishing their images, and instantiate / alter it later.
Once the \EXPDATE is reached, the images become unaccessible.  

\subsection{Protocol Overview}
\begin{figure}[t]
  \begin{center}
    \includegraphics[width=0.85\linewidth]{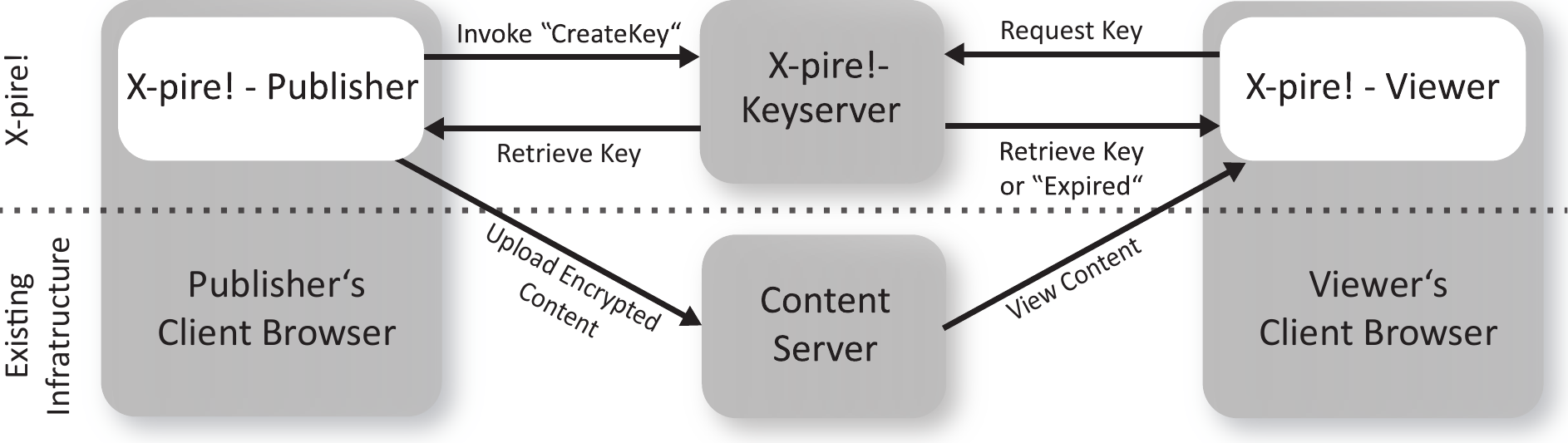}  
    \caption{Schematic overview of our approach}
    \label{fig:overview}
  \end{center}
\end{figure}

A high-level overview of the protocol and the involved participants is
given in Figure~\ref{fig:overview}. The involved participants of the protocol are the \emph{content
  publisher} $\mathcal{P}$, the \emph{content server} $\mathcal{C}$ (which is often a
social network), and the \emph{viewer} $\mathcal{V}$.  We add a specialized
\emph{\KEYSERVERA} $\mathcal{K}$ to the setup, who is trusted not to hand out
keys after their expiration date has been reached.

Our approach distinguishes three phases: 1) the content
\emph{publishing phase}, i.e., adding the \EXPDATE and storing the
resulting images, 2) the \emph{viewing phase}, i.e., visiting the
corresponding websites containing these images, and 3) the
\emph{update phase}, i.e., shortening or prolonging the
\EXPDATE of published images.

\subsubsection{Publishing Phase}
When \emph{publishing} an image $d$, the publisher $\mathcal{P}$ first contacts the
\KEYSERVERA $\mathcal{K}$ with the request for a key.  The \KEYSERVERA
creates a new key $k$ and returns it to the \PUBLISHERA.  $\mathcal{P}$ now encrypts the
image $c \leftarrow E_{k} (d)$, hashes this ciphertext $hash \leftarrow
H(c)$, and sends the hash $hash$ and the \EXPDATE $expdate$ back to the keyserver,
which adds the hash and the \EXPDATE to the respective entry.
$\mathcal{K}$ finishes the key creation by sending the key ID $id_k$
back to the \PUBLISHERA. The encrypted data $c$ is embedded into a valid JPEG image (see
Section~\ref{sec:embedding_encryptions_to_JPEG_intro}) and stored on
the content server $\mathcal{C}$.

The reason to store keys rather than encrypted images is that keys are
much easier to handle due to their small size.  Thus users can utilize
the large storage capacity offered by the websites/social networks,
and avoid turning the keyserver into a performance bottleneck.

\subsubsection{Viewing Phase}
When a user wants to \emph{view} an encrypted image $c$ on a website,
the user sends its hash $H(c)$ and the key identifier $id_k$ of the
key $k$ retrieved from the image to the \KEYSERVERA. The \KEYSERVERA
checks that the \EXPDATE has not yet been reached, checks the
correctness of the additional data, and (if all these checks passed
successfully) sends back the key $k$. If the key gets delivered, i.e.,
the \EXPDATE of the key $k$ has not yet been reached, the image $c$
decrypted, and the user is able to \emph{view} the image $d$. The
reason to send the hash $H(c)$ is to provide convincing evidence that
the corresponding image was really downloaded. This renders a simply
key-requesting attack much more costly in terms of band-width
consumption.

\subsubsection{Update Phase}
We provide a comprehensive key-management interface that allows users
to later review all previously generated keys along with a description
of the corresponding images. This enables the \PUBLISHERA to
\emph{update} existing expiration dates. In particular, a publisher can
prolong or shorten existing expiration dates, and even enforce
instantaneous expiration of published images if desired.
We describe the protocol in more detail in
Section~\ref{sec:tech_details}.

\subsection{Embedding encryptions to JPEG}
\label{sec:embedding_encryptions_to_JPEG_intro}
Our approach requires the encryption data $c$ to be stored on the
content-server $\mathcal{C}$.  This task is inherently more difficult
for images, since social networks such as Facebook and Flickr scale
and re-encode images before storing them (in contrast to scenarios
where data is not subsequently manipulated
\cite{geambasu-09-vanish,ephemeral-data-project-2010}).  First, this
means that an upper bound on file size and image dimensions is
enforced upon uploaded images. Second, and more important, the image
is re-encoded using JPEG compression. Thus, simply uploading the
encrypted data $c$ would result in the publication of a re-encoded
version $c'$, in which the encryption would be fully destroyed. As a
consequence, nobody would be able to decrypt the image anymore; it
would expire instantaneously. Therefore, a robust embedding is
required.  Common steganographic techniques do not achieve
sufficiently good data rates to robustly embed high quality images in
images facing such constraints.
We have developed a novel, robust embedding of arbitrary data into
JPEG images that achieves sufficiently good data rates such that
encrypted images can be embedded into JPEG images that are accepted by
social networks.  We describe this embedding in detail in
Section~\ref{sec:jpeg_embedding}.

\subsection{On the centralized approach}
\label{sec:on_the_centralized_approach}
Decentralized solutions are often considered superior to centralized
ones, as they typically offer better performance and scalability.  In
the situation we are facing in this paper, however, we believe that a
centralized approach has advantages that outweight the benefits of
decentralization, as described in detail below. We hence decided to
ground \XPIRE on a centralized keyserver, but we stress that our
technique for dealing with post-processing of images, which arguably
constitutes our main contribution, can be similarly incorporated to
decentralized systems like Vanish~\cite{geambasu-09-vanish}.

\begin{compactitem}
\item Using a centralized keyserver allows us to increase the costs for an attacker to retrieve keys by
  requesting users to provide convincing evidence that they actually
  know the data for which they are requesting the key for. This is
  ensured by storing and matching the hash of the encrypted data.
\item Using a central keyserver we can limit the rate of key requests
  possible from single IP-addresses and IP-ranges.
\item If desired, our approach enables the \PUBLISHERA to additionally
  require users to solve \CAPTCHAS before viewing the image (e.g. one
  for every photo album they are viewing). This can be easily realized
  by having the centralized keyserver generate the \CAPTCHA,
  and it effectively raises the bar that a large number of
  keys will be downloaded using automated crawling.
\end{compactitem}
We describe more details about the measures taken against collecting
keys in Section~\ref{sec:tech_details}.

Besides these security issues, a central \KEYSERVERA also provides
more flexibility in the expiration dates than decentralized solution
attempts. In addition to providing faster access, a centralized
approach exceeds a peer-to-peer setting in terms of scalability in
this scenario, where only short keys are stored.
Moreover, it is arguably more privacy-friendly to trust a dedicated
keyserver rather than trusting the social networks, whose business
model relies crucially on the collection and openly distribution of
user data.  A centralized keyserver does not have a business incentive
to act unexpectedly, but rather would lose its credibility if it
becomes public that the server acts inappropriately with these
keys. We currently provide our own dedicated \KEYSERVERA, but we
additionally released the \KEYSERVERA application such that other
trustworthy authorities or individuals can set up a keyserver as
well. In the case that a user is not willing to trust any of these
keyservers, he can still set up a personal
\KEYSERVERA himself to avoid such trust dependencies.

\subsection{Implementation}
We provide a complete working implementation of the
protocol.
The Browser extensions are implemented as a Firefox plug-in that is
functional on all major platforms (Windows, Linux, Mac).  The image
embedding part as described above is currently implemented to work on
Facebook~\cite{Facebook}, Flickr~\cite{Flickr}, and
wer-kennt-wen~\cite{WKW} (a famous German social network). We do not
expect major problems in adding support for further social networks,
since all social networks we are aware of rely on common techniques
for JPEG compression based on the libJPEG.
To store an image, the plug-in prompts for the files, and performs all
operations described above.  It stores the processed files in a
temporary directory, from where they can be uploaded by any
upload-mechanism that is offered by the social network.

When users are visiting a website, the plug-in searches for protected
images, automatically retrieves the required keys from the \KEYSERVERA,
and displays the decrypted image inline on the page without any
further user interaction.
If the plug-in is not installed, these images show a brief statement
that the required plug-in is missing, along with a link where the
plug-in can be downloaded.
A detailed description of the process of detecting and displaying
protected images can be found in Section \ref{sec:implementation}.

\section{Detailed Protocol Description}
\label{sec:tech_details}
In this section, we give a detailed description of the protocol that
underlies our approach. The protocol considers three participants:
the data \PUBLISHERA $\mathcal{P}$, the data \VIEWERA $\mathcal{V}$,
and the \KEYSERVERA $\mathcal{K}$.\footnote{We sometimes additionally
speak about a content server $\mathcal{C}$: an arbitrary server of one
of the supported social networks that takes care of storing and
providing the content that is created by our protocol. However, this
entity $\mathcal{C}$ is not required to take part in the actual
protocol.}
Consequently, our software is split into three applications: the
\PUBLISHER, the \VIEWER, and the \KEYSERVER. The \PUBLISHER
establishes a communication channel between the publisher and the
keyserver, and adds an \EXPDATE to the desired images before they are
uploaded to, e.g., Facebook, in the usual manner. The \VIEWER permits
the viewers to display these uploaded images, provided that the
respective expiration dates have not yet been reached. In order to
display these images, the \VIEWER establishes a communication channel
between the viewer and the keyserver. Finally, the \KEYSERVER creates,
stores, and hands out the necessary keys as described above, and it
allows publishers to subsequently update the expiration dates of their
keys.
 
In the following, we describe these communication protocols in detail. 
We distinguish three phases:
the content \emph{publishing phase}, the \emph{viewing phase}, and the 
\emph{update phase}.

\subsection{Publishing Phase}
In the publishing phase, the \PUBLISHERA requests a key by invoking a
key creation process on the keyserver (\textit{CreateKey})
and by providing his account
information $cred$. The \KEYSERVERA generates the key $k$ and sends it
along with a session ID $id_s$ back to
the \PUBLISHERA (see below for how $id_s$ is used). The \PUBLISHERA
receives the key and encrypts the image $d$ that he wants to upload with
$k$ as 
\[
c \leftarrow E_k(d).
\]
For encryption we use the Advanced Encryption Standard
(AES)~\cite{AES} with a key length of 256 bits and Cipherblock
Chaining Mode (CBC)~\cite{CBC}. 

The ciphertext $c$ needs to be
embedded into another image file $container$ to obtain a valid image
file $emb$
(details of the embedding are shown in
Section~\ref{sec:wheretoplaceencinjpeg}):
\[
emb \leftarrow \textnormal{Embed}(container,c)
\]
Without embedding $c$ into
$emb$ it would not be possible to upload it as an image to a
social network since $c$ constitutes a ciphertext and hence does not 
fit the format of an image file.

We stress that developing a suitable embedding is difficult, since
social networks commonly apply image post-processing techniques, and
thus to be suitable, an embedding must be resistant to JPEG
recompression.
 
After encrypting the image $d$ to the ciphertext $c$, the \PUBLISHERA
replies by sending the hash of the newly encrypted image $c$ back to
the keyserver. The hashes are computed by applying the
SHA256~\cite{SHA256} hash function $H$ to the encrypted images:
\[
hash \leftarrow H(c).
\]
The hashes are sent to the keyserver by invoking \textit{AddHashes}
on the server side using the previously received session ID $id_s$
and the following additional payload:

\begin{compactenum}
  \item $expdate$: the expiration date chosen by the user,
  \item $hash$: the hash of $c$, and
  \item $description$: an identifier to describe (collections of) images; 
        it has to be explicitly set by the user.

\end{compactenum}

In order to complete the publishing phase, the \KEYSERVERA stores the
key $k$, the \EXPDATE $expdate$, the description $description$ and the
hash $hash$ to its database thereby creating the unique key ID $id_k$,
which is also stored. Finally, it acknowledges the key creation
process by returning $id_k$ of the previously created key $k$ to the
\PUBLISHERA.

The $hash$ is required because it serves as convincing evidence that the whole
data has been downloaded before the key was requested.  Finally, the
identifier $\mathit{description}$ plays two important roles in our protocol.

First, it is used for a subsequent \emph{key management}. Together
with the creation date and the expiration date, the description yields
a unique identifier that later enables users to identify specific
encryption keys, e.g., for prolonging or shortening the expiration
date. In particular, by changing the \EXPDATE it is possible to let
images expire instantaneously if desired. Furthermore, all images
prepared in the same step using one $\mathit{description}$ share the
same expiration date.

Second, it allows for \emph{grouping images}, and for a controlled
\emph{use of Captchas}.
If several images are prepared for uploading at the same time using
the same $\mathit{description}$, all images are encrypted with the
same key. This allows for syntactically grouping images, and for a
simpler decryption functionality, where only one key needs to be
requested for one set of images.
Moreover, it allows for a controlled usages of \CAPTCHAS: The
publisher can decide to require \CAPTCHAS for additional security.
In this case, every viewer will have to solve a \CAPTCHA before the
correct image is displayed. Images with the same description will
receive one joint \CAPTCHA.  The idea behind a set of images using the
same $\mathit{description}$ and therefore having the same encryption
key is that, e.g., whole albums can be viewed after providing the
solution to a single \CAPTCHA. From a security perspective, disabling
\CAPTCHAS, or only asking for one \CAPTCHA per collection of images,
reduces the security and increases the vulnerability to automated
crawling attacks.  However, for reasons of efficiency and seamless
integration, we prefer to keep the number of \CAPTCHAS low.

\subsection{Viewing phase}
After the image $emb$ with the embedded ciphertext $c$ has been created, it can be 
uploaded to the target web site, e.g., to Facebook. From that time onward, it can be viewed using the
\VIEWERA plug-in by any eligible user that has access to
the (encrypted) image on Facebook.
The \VIEWERA is required to decrypt the prepared images and to
correctly display them, provided that they have not yet reached their
expiration date. Without the \VIEWERA, only the container images
providing a link to the \VIEWERA application are shown; the actual
image stays encrypted.

When a user enters the website of one of the supported social networks 
with a browser that has the
\VIEWERA application installed, the \VIEWERA starts to search for
encrypted images. If it detects an encrypted image, it starts the
decryption process. The \VIEWERA extracts the key identifier $id_k$ as
well as the address of the server storing the key ($keyserver$) and
computes the SHA256-hash of the encrypted image. This information is
used to invoke the \textit{GetKey} process at the \KEYSERVERA. The
server performs a key lookup, and distinguishes two cases: If the
expiration date has already been reached, it returns an error code
indicating the expiry; if the expiration date has not yet been
reached, the server checks whether the settings of the key require the
\VIEWERA to solve a
\CAPTCHA to see the decrypted picture.

If no \CAPTCHA is required, the server sends the requested key
along with $id_s$ back to the \VIEWERA.
The viewer uses the retrieved
key to decrypt the detected image.  Afterwards, the detected image is
locally replaced by the decrypted one.

If a \CAPTCHA is required, the server sends a \emph{CAPTCHA\_REQUIRED}
request to the \VIEWERA. 
The \VIEWERA requests a
challenge from the \CAPTCHA service using the public key of the
\KEYSERVERA, the user solves this challenge
and sends it together with the solution to the \KEYSERVERA.  
The \KEYSERVERA uses its private key and
verifies together with the \CAPTCHA service whether the solution was
correct.  
If a
wrong solution for the \CAPTCHA is sent to the
\KEYSERVERA, the server notifies the \VIEWERA that the solution was wrong and
invalidates the challenge during its verification with the \CAPTCHA
service. The \VIEWERA then decides whether to give up or try again by
requesting a new \CAPTCHA. If the correct solution is provided, the
server sends the requested key along with $id_s$ to the \VIEWERA as
described above.  
Adding
\CAPTCHAS  results in additional communication between the \CAPTCHA
service and the \VIEWERA as well as between the \CAPTCHA service and
the \KEYSERVERA.

\subsection{Update Phase}
Our protocol allows users to manage and update the keys
that they have already created with the \KEYSERVERA. The
\KEYSERVERA offers a web interface that provides all necessary
functionality to identify already existing keys using a unique
identifier comprised of the $description$, the create date, and the
expiration date, and to modify the current settings. The main
functionality of the key management is to allow the \PUBLISHERA to
prolong or shorten the expiration dates for keys that have already
been created. In particular, it allows a user to enforce an
instantaneous expiration of his  images.

\subsection{The Embedding}
\label{sec:wheretoplaceencinjpeg}
The main goal of our approach is to find a method for adding an
expiration date to an image that is subsequently hosted on an existing
web site. Because encrypting images and later restricting the key
access while retaining a valid image files is the only possibility to
solve this problem, we must embed all data needed for the decryption
into the image file itself. To publish images to the internet users
can either provide them inside of plain HTML-pages by the $\langle
img\rangle$-tag (users need to have access to the source of these
pages in order to add images) or provide them within special Internet
services like social networks (e.g. Facebook or Flickr) where users
provide their images as image files to an upload-interface.  To the
best of our knowledge, although social networks support
several image formats for uploading, all existing social networks
store the final image as a JPEG file. Therefore, our approach provides
a solution for JPEG files. Since the expiration date is implemented by
encrypting data and restricting the access to the keys needed for
decryption to the time until the expiration date is reached, all
information needed to retrieve the key for decrypting the image needs
to be stored inside of the JPEG file:

\begin{compactitem}
	\item{the keyserver address $keyserver$ to place the query,}
	\item{the key ID ${id_k}$ to identify the needed key,}
        \item{the version ID $version$, and}
	\item{the encrypted image $c$ to compute the hash
            $hash$ and to decrypt the image itself later.}
\end{compactitem} 

We have to decide where in a valid JPEG file the data can be embedded,
without causing the file to be rejected by upload-interfaces. Inside
of the $\langle img\rangle$-tag, HTML expects one of the supported
image formats (e.g. JPEG, PNG, GIF). If a JPEG file is encrypted by
standard methods, the outcome is a ciphertext and not an image file.
Therefore, it is necessary to embed the ciphertext into another
``container''-JPEG such that the outcome is still a valid JPEG. In
general, one can think of two approaches to achieve this: one either
stores the encryption within a \emph{header field}, or \emph{inside
  the actual image data}. We will discuss both approaches in detail in
Section~\ref{sec:embeddingworkflow} after we have introduced relevant
background information on JPEG in Section~\ref{sec:jpeg_details}, but
we anticipate the major lessons learned and results here: Embedding
the encryption in the header information is significantly more
efficient and easier to handle; however, virtually all social networks
remove all header information for uploaded images, rendering this
approach completely useless for those sites. The alternative --
embedding within the actual image data -- is significantly more
challenging because our embedded encryption must survive the JPEG
recompression that these sites apply to the image data. We developed
a suitable such embedding that renders the approach possible and
practical for social networks.

\section{Surviving JPEG Compression}
\label{sec:jpeg_embedding}

\begin{figure*}[t]
  \begin{center}
    \includegraphics[width=1.0\linewidth]{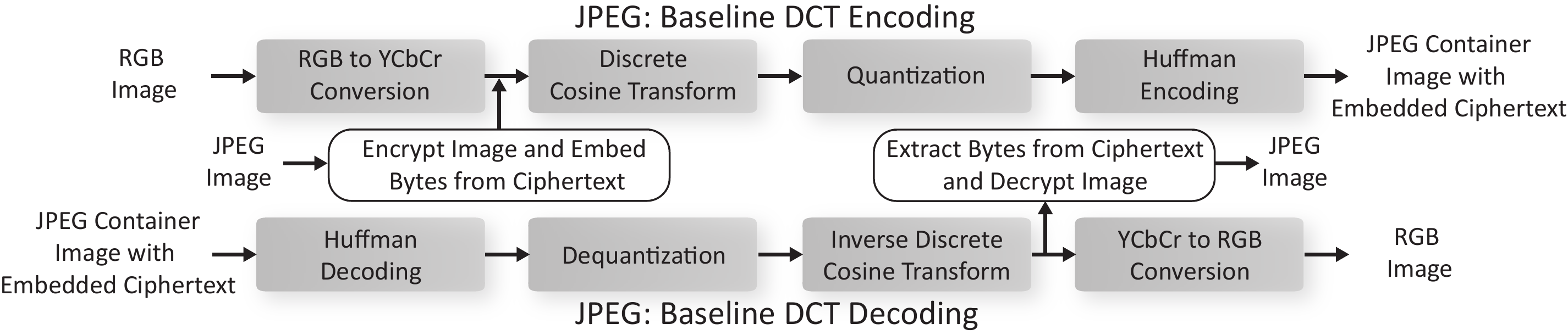}
    \caption{Overview of JPEG encoding and decoding}
    \label{fig:jpegenc}
  \end{center}
\end{figure*}

In this section we first provide general background information about
the JPEG standard. Afterwards we discuss in detail the two embedding
approaches mentioned above.

\subsection{Details on JPEG}
\label{sec:jpeg_details}
The JPEG standard was developed by the Joint Photographic Experts
Group in 1992~\cite{jpeg_standard} and is the current de-facto
standard for compressing images in the Internet. A JPEG image is most
commonly stored in a file according to the JPEG File Interchange Format
(JFIF)~\cite{JFIF}. This standard defines the complete structure
of the file and describes how and where to store all the information
(e.g. Huffman tables, quantization tables) needed to later decode the
image again.  Usually, when people refer to JPEG files, they are actually 
referring to JFIF files.

We focus on the \emph{Baseline DCT} (BDCT) mode of JPEG in our
description, since it is the most widely used mode in the Internet,
and also the mode used in the social networks currently supported by
\XPIRE. In the following, we describe the individual steps of the JPEG
encoding; an overview is given in Figure~\ref{fig:jpegenc}.

\subsubsection{RGB to YCbCr Conversion}
The starting point for creating a BDCT image is an RGB
image~\cite{pennebaker_jpeg} consisting of $x$ by $y$ pixels. RGB
images store for every pixel $(x,y)$ the red, green and blue color
components. All three color channels together are converted to the
YCbCr~\cite{pennebaker_jpeg} color model that consists of the
luminance channel Y and the two chrominance channels Cb and Cr. The
luminance channel Y provides information about the brightness, whereas
the two chrominance channels Cb and Cr provide color information. The
conversion between the two color models is not necessarily lossy.
However, divisions often introduce rounding errors when real valued
results are converted back to integer values, and both the input
RGB-values and the resulting YCbCr-values are typically integer
values.

As an optional step, the color model conversion might be followed by
subsampling. Subsampling was not included in the original JPEG
standard, but was later published as an extension
\cite{jpeg_standard84}.  Subsampling reduces the size of a channel by
downsampling it in a horizontal and/or vertical direction. Subsampling
is typically applied only to the chroma channels; we hence assume here
that the luminance channel is left unchanged. A common case for highly
compressed JPEG images is a 2x2 chroma subsampling where only half of
the information in the horizontal and vertical directions is kept.
Thus, if we have an image size of $x$ by $y$ pixels (which is also the
size of the luminance channel), the chroma channels each have only
$x/2$ by $y/2$ pixels, and thus they are only $1/4$ of their original
size. We refer to ~\cite{jpeg_snoop,pennebaker_jpeg,jpeg_standard84}
for more detailed information about subsampling.

After the color model conversion and the optional subsampling, the
image is level-shifted by subtracting 128 from all values. This
produces a signed result with smaller values. Next, the image is split
up into blocks. Each block consists of 8 by 8 pixels (default size for
BDCT) and is extracted from the image as shown in
Figure~\ref{fig:jpegblocks} in Appendix~\ref{sec:append_JPEG_Exp}. The
image block's values are called the DC-component (the value at
position (0,0)) and AC-components (all other values).

\subsubsection{Discrete Cosine Transform}
The extracted blocks are used as input for a discrete cosine transform
(DCT). 
The DCT function gets as input an
image block with integer values and outputs a block with real values.

\subsubsection{Quantization}
Quantization plays a crucial role for the compression of a JPEG image
and is performed by dividing an image block by the so-called
quantization table. A quantization table consists of 64 values, and
each value corresponds to one value inside an image block (see Figure
\ref{fig:fbookqtable} in Appendix~\ref{sec:append_JPEG_Exp} for a
quantization table used by Facebook).  The division is done by
dividing a value from the image by its corresponding value from the
quantization table and rounding the result. The values of the
quantization table define the compression rate and can be
user-defined.  Every single value has to be in the interval from 1 to
255.  If all values inside the quantization table are 1, no
compression is in place; the higher the values, the higher the
compression rate. The quantization tables are stored inside of the
JFIF-file and used again later during the decoding of the
image.

\subsubsection{Huffman Encoding}
Finally, the DC coefficients (the values at position $(0,0)$ of each
block) are difference-encoded, a zig-zag reordering is applied, and
the result is Huffman-encoded. The difference encoding works as
follows: every DC coefficient is replaced by the result of
substracting the previous blocks' DC coefficient from the current DC
coefficient.  Although Huffman encoding optimizes the final image
size, the encoding is lossless, similarly for the difference encoding
and the reordering. The information needed for a later Huffman
decoding is stored inside of the so-called Huffman tables; they differ
for the DC and AC components and for the luminance and chrominance
channels. All four tables are stored analogously to the quantization
tables, i.e., they are stored inside of JFIF files for a later
decoding.

In order to decode a JPEG image, the procedure is simply reversed.
Starting with the JPEG file, a Huffman decoding is applied first. Then
the reordering is removed which is followed by undoing the difference
encoding.  A dequantization is then applied using the same
quantization matrices that were used during the encoding. After the
inverse discrete cosine transform has been applied, the values are
shifted back (undoing the level shifts from the encoding) and finally
the YCbCr color channels are converted to the RGB colorspace.

\subsection{Robustly embedding encryption into JPEG}
\label{sec:embeddingworkflow}
We have pursued two main approaches for including the ciphertext in
the JPEG image. The straightforward way is to embed the encrypted
image into a special header field of a JFIF file; the alternative
approach is to embed the encryption into the actual image data of a
JFIF file.

Whenever the websites that host the image retain header information,
placing the encryption in the header is clearly the best method in
terms of performance. However, this approach is not feasible for
common social networks, since they immediately remove all header
fields contained in images. For social networks, we thus have to embed
the ciphertext in the image data of the JFIF file. However, when
images are uploaded to a website via a web-interface, such as for social
networks, they are often recompressed. This recompression is highly
problematic for us: By the cryptographic properties of secure
encryption schemes, we will only be able to decrypt the encrypted
image if it is recovered with 100\% accuracy (bitwise). Otherwise the
decryption will fail, or provide non-sensical results.  In order to
achieve the goal of 100\% accuracy, we have to embed the encryption
into a JFIF file so that the embedding survives the JPEG
recompression. In the following, we present and discuss both
approaches.

\subsubsection{Embedding encryption inside of header fields}
The JPEG standard supports comments inside the image header, which can
be also used for storing the encrypted image. The length of each
comment field is limited, but the number of comment fields inside the
header is unconstrained.

This solution solves the problem for web applications where images are
provided for viewing without any further modifications. The classic
example is a static website where images are stored on the webserver
and linked using the $\langle img\rangle$-tag of the HTML standard.
We have implemented the embedding into headers fields as one of the
supported solutions for these classic websites. For these websites,
placing the encryption in the header is clearly the best method in
terms of performance.

\subsubsection{Embedding encryption inside of image data}
When images are uploaded to social networks they are often scaled and
recompressed. The problem of scaling is solved in our approach by
providing encrypted images inside a container image that exactly
matches the sizes expected by social networks so that rescaling does
not occur. We describe in the following how we address the problem of image
recompression. 

Embedding an encrypted image into the image data in a way that it
survives JPEG recompression turned out to be a challenging task.  For
the sake of exposition we first describe an intuitive approach for
embedding the encrypted image inside of image data that we initially
pursued, and then described where it fails. After that, we explain the
modifications we had to make to actually solve the problem.

The first approach we pursued when we recognized that social networks
decompress JPEG images and then compress them again with their own
settings (compression ratio, header fields etc.), was to compute a
preimage for the JPEG image that we wanted to upload to the social
network, i.e., the image that contains the encryption. The underlying
idea was to embed the data into a container image and to take this container image with the embedded data as the
starting point. Let us call it $\mathit{Image}$.  Knowing that
$\mathit{Image}$ is exactly the image we want to have on the website
later, it is necessary to go through the compression steps done by the
social network in the opposite direction. We want to know precisely
which $\mathit{Image'}$ we have to provide for the upload such that
after the compression the outcome on the website is exactly
$\mathit{Image}$. To achieve this, all functions applied during the
compression have to be inverted step by step. After we compute
$\mathit{Image'}$, we must provide an $\mathit{Image''}$ without any
JPEG compression to the social network, such that decoding it results
exactly in the computed preimage $\mathit{Image'}$. However, going
this direction turned out to be infeasible because this approach
includes inverting the Huffman-encoding without having the
Huffman-tables. Without the tables, the information needed to decode
the image is not available and therefore the decoding can only by done
by brute force. 

Our second approach was to modify our input image such that the
compression step using the quantization table cancels out. This would
provide us with a resulting image in which we can control the content of
the luminance channel. To achieve a canceling out of the quantization,
its inverse needs to be applied beforehand. Therefore, we modify the
quantization table used for encoding the JPEG image that is initially
uploaded to the social network. This causes the dequantization
done during the image decoding inside of the upload routine to exactly
apply the inverse of the quantization done later during the encoding.
Let $A$ be the quantization table used by the social network, then we
can use a quantization table $A'$ for the initial image that is
achieved by a point-wise division with $A$ as follows:

\begin{eqnarray}
	A=(a_{i,j}), A'\hspace{-.5mm}=(a'_{i,j})=\left(\frac{1}{a_{i,j}}\right) \hspace{-.5mm}\textnormal{for } i,j\in\{0,\dots,7\} \hspace{-2mm}\nonumber
\end{eqnarray}

This solves the problem of embedding encryptions into JPEG images such that
they survive the JPEG recompression, at least in theory: This approach
worked perfectly well with our implementation, which strictly follows
the JPEG standard.

\subsection{Dealing with libJPEG and its idiosyncrasies -- how to achieve robustness in practice}
However, when using the libJPEG implementation\footnote{libJPEG is a
popular and widely used implementation of the JPEG standard, written
and maintained by the Independent JPEG Group~\cite{IJG}. All social
networks we are aware of use libJPEG directly, or use image
manipulation programs that rely on libJPEG.} this approach broke down
completely. The computations done in libJPEG are highly optimized and
introduce, for example, rounding errors that our mathematical accurate
computations according to the JPEG standard did not have. In general,
a small amount of loss is not really a problem for images since users
do not detect such small differences, but in our case we need enough
accuracy to reconstruct the embedded data 100\% correctly to perform
the subsequent decryption. In this setting our reconstruction rate was
only 10\%, which made it impossible to amplify this approach using
error-correcting codes. Since the social networks our software targets
use libJPEG, the limited accuracy of libJPEG also rendered our second
approach infeasible.

\medskip\noindent\textbf{libJPEG-Details.}
Before we analyze how to circumvent the accuracy problems introduced
by libJPEG, we will have a closer look at what libJPEG does.
The library is entirely implemented in C. A typical
workflow for an application using libJPEG for image manipulation is as
follows: let libJPEG decompress the source image, define parameters
for the resulting image, and let libJPEG compress the image again
using the defined parameters. Performing only a specific operation of
the JPEG compression process, e.g., the quantization in Figure
\ref{fig:jpegenc}, by just calling a ``quantize'' method is not
possible, as the library combines steps of the process for optimized
performance. Moreover, operations like scaling are combined with the
DCT computation, at some places precision is sacrificed in favor of
speed, and many special cases of the same operation are handled by
individual functions. As a result, using libJPEG as intended, i.e., to
perform a whole image compression process, is easy, whereas performing
JPEG compression steps one by one as shown in Figure \ref{fig:jpegenc}
is close to impossible.

All social networks in the Internet we are aware of either use libJPEG
directly, or use image manipulation programs like GD~\cite{GD}, and
ImageMagick~\cite{ImageMagick} that rely on libJPEG. To the best of
our knowledge, all of them follow the intended workflow. They just call
the library to perform the intended operations without doing any
in-depth manipulation.

\begin{figure*}[t]
  \begin{center}
    \includegraphics[width=0.9\linewidth]{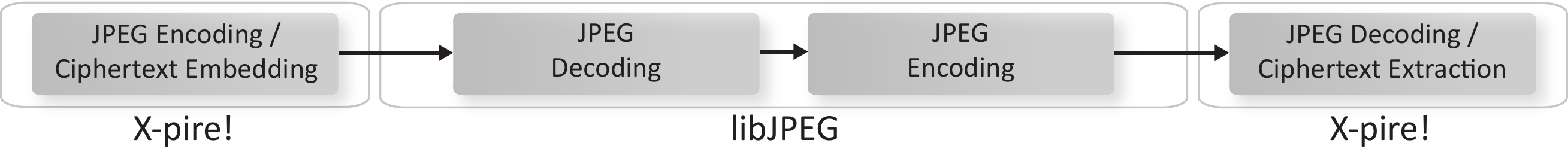}
    \caption{Workflow for embedding encryption into JPEG}
    \label{fig:embedworkflow}
  \end{center}
\end{figure*}

For our approach, however, it is crucial that we can interfere with the
JPEG compression and decompression processes. This requires an
in-depth understanding of the libJPEG functions and increases the work
needed to embed the necessary functionality.

\medskip\noindent\textbf{Dealing with libJPEG.}
Since libJPEG's inaccuracy rendered our mathematically-correct
approach infeasible, it was necessary to have an approach for
embedding data that survives the compression done by the quantization
step. Since the quantization tables used for the compression are
publicly known -- usually they do not change for one service and can
be extracted from every JPEG image --, we must try to embed data such
that applying the division inside the quantization does not destroy
the embedded information.

If we have a look at the quantization table used for example by
Facebook, we can see that the highest value used in the luminance
table is 36. The values inside of one image block that are divided by
the quantization values are 8-bit unsigned integers. If we look at the
binary representation of such a value, a division by 36 does not
destroy the two most significant bits. Therefore, we embed two bits as
the two most significant bits in every value of every block inside of
the luminance channel. In order to improve the results and to gain
more stability, we also always set the fifth bit to one and bits
zero to four to zero. This protects bits six and seven from being
affected by rounding errors, as it prevents even significant
deviations in the lower bits from reaching the upper bits. 

Embedding two bits works as long as the divisor needs at most six bits
in binary representation. For all social networks of which we are
aware, the maximal divisor has at most six bits. In general, when
quantization tables with higher values are used, it might only be
possible to embed one bit or even none. If smaller values are used, it
would be possible to embed more than two bits per byte. For the
future, one could use the knowledge of the quantization tables to
design an approach that dynamically optimizes the number of bits to be
embedded per coefficient. With the current approach of embedding two
bits per byte we have an upper bound on the amount of bits we can
embed: $1/4$ of the original luminance data, two bits per pixel of the
luminance channel instead of eight. If subsampling is used for the
luminance channel, the maximum amount that can be embedded decreases
accordingly. However, the amount of information that we can embed
using our technique is more than sufficient for embedding encrypted
images into cover images that are permitted by the social networks,
and thus for realizing the desired expiration date. 
The white boxes in Figure~\ref{fig:jpegenc} show at which point of the
JPEG encoding and decoding processes the ciphertext bits are
embedded and extracted.

Note that although we have eliminated most loss by embedding the
encrypted data only in the two most significant bits, a small amount
of loss can still occur, e.g., because of internal rounding mistakes.
Thus, it is necessary to apply an error-correction method. The current
approach uses Reed-Solomon
codes~\cite{reedsolomon-1960-polynomial-codes} with a configuration
$(N,K)$ of $(255, 191)$.  This means one error correction unit
consists of 255 bytes, 191 bytes data and 64 bytes of additional
information. The introduced overhead reduces the real space for
embedding information to $3/16$ of the original payload since the $64$
parity bytes per block have to be embedded as well.  Using this
configuration it is possible to recover the correct ciphertext even if
up to 12.5\% of each error correction unit is lost.\footnote{We
  never encountered more than 5\% incorrect bits in our
  experiments with this approach, i.e., a significantly weaker
  error-correcting code with a smaller overhead would have been
  sufficient. However, we decided to deploy a stronger code to catch
  unexpectedly bad results that might arise. For the same reason, we
  currently only embed two bits in every value of every block, even
  though our experiments indicated that more bits could have been
  embedded in most cases.} 
The complete workflow starting with
embedding bits until the final extraction of the encrypted image is
shown in Figure~\ref{fig:embedworkflow}.

We recompress our images before we embed them with quality settings
similar to the ones used in social networks. Consequently, although we
embed data only into the two most significat bits per byte, our method
can be applied to achieve essentially the same image quality in
practice as images that have been uploaded directly to social networks
like Facebook. We refer to Appendix \ref{sec:appendix:imagequal}
for bounds on possible image sizes and illustrative images.

\section{Implementation}
\label{sec:implementation}
We have implemented the publisher and the
viewer as Mozilla Firefox extensions, so that the user can
prepare protected images when browsing his social network site without
switching programs. The extraction of protected images is completely
automated and opaque to the user.

Both the publisher and the viewer application share a core library 
that provides encryption, JPEG manipulation, and error correction,
as well as a high-level communication functionality.

\begin{figure}[ht]
  \begin{center}
\begin{minipage}[t]{0.32\textwidth}
    \begin{tikzpicture}
      \begin{axis}[
        xlabel=Number of pictures,
        ylabel=Creation time in seconds,
        width=1.0\textwidth,
        height=5cm,
        enlargelimits=false,
        ytick={0,1,2,3,4,5},
        yticklabels={0,1,2,3,4,5},
		grid=major, 
        scaled ticks=false, 
		legend style={at={(0.5,+1.1)},anchor=south,cells={anchor=west}}]

        \addplot[color=deepred]
        table[x=pictures,y=creationtime]{benchmark-bits.data};
		\addlegendentry{\footnotesize{bit embedding}}
        \addplot[color=lightorange, mark=*]
        table[x=pictures,y=creationtime]{benchmark-header.data};
		\addlegendentry{\footnotesize{header storing}}
      \end{axis}
    \end{tikzpicture}
    \caption{Creation time 
}
    \label{fig:perf_encrypt}
\end{minipage}
\hfill
\begin{minipage}[t]{0.32\textwidth}
    \begin{tikzpicture}
      \begin{axis}[
        xlabel=Number of pictures,
        ylabel=Seconds to display the website,
        width=1.0\textwidth,
        height=5cm,
        enlargelimits=false,
        ytick={0,0.1,0.2,0.3,0.4,0.5},
        yticklabels={0,0.1,0.2,0.3,0.4,0.5},
		grid=major, 
        scaled ticks=false, 
		legend style={at={(0.5,+1.1)},anchor=south,cells={anchor=west}}]
        
        \addplot[color=deepred]
        table[x=pictures,y=time]{benchmark-withoutplugin.data};
		\addlegendentry{\footnotesize{without plugin}}
        \addplot[color=lightorange, mark=*]
        table[x=pictures,y=time]{benchmark-noxpire.data};
		\addlegendentry{\footnotesize{with plugin,}}
      \end{axis}
    \end{tikzpicture}
    \caption{Time to display websites}
   \label{fig:perf_view}
\end{minipage}
\hfill
\begin{minipage}[t]{0.32\textwidth}
    \begin{tikzpicture}
      \begin{axis}[
        xlabel=Number of pictures,
        ylabel=Seconds to display the website,
        width=1.0\textwidth,
        height=5cm,
        enlargelimits=false,
		grid=major, 
        scaled ticks=false, 
		legend style={at={(0.5,+1.1)},anchor=south,cells={anchor=west}}]legend pos=north west]
        
        \addplot[color=deepred, mark=diamond*]
        table[x=pictures,y=time]{benchmark-invalidcomments.data};
		\addlegendentry{\footnotesize{header storing, invalid}}
        \addplot[color=lightorange, mark=*]
        table[x=pictures,y=time]{benchmark-validcomments.data};
		\addlegendentry{\footnotesize{header storing, valid}}
        \addplot[color=deeppurple, mark=x]
        table[x=pictures,y=time]{benchmark-invalidbits.data};
		\addlegendentry{\footnotesize{bit embedding, invalid}}
        \addplot[color=deepblue, mark=triangle*]
        table[x=pictures,y=time]{benchmark-validbits.data};
		\addlegendentry{\footnotesize{bit embedding, valid}}
      \end{axis}
    \end{tikzpicture}
    \caption{Time to decrypt images}
    \label{fig:perf_decrypt}
\end{minipage}
  \end{center}
\end{figure}

The core library used by both applications is responsible for the
actual handling of protected images, i.e., modifying JPEGs, encryption
and decryption, computing hashes, and error correction. It is
implemented in C and C++. The reason for the separation of core
library and client applications is the availability of fast and
well-tested libraries for cryptography, JPEG manipulation, and
error-correcting codes in C, as well as the lack thereof in
Javascript. However, Javascript is much better suited for high-level
server communication than C or C++ and is therefore used for the
communication task. In particular, we rely on OpenSSL for hashes and 
encryption, libJPEG
for JPEG manipulation, and our own implementation of Reed-Solomon
codes for error correction. For embedding data into JPEGs for
social networks, we extended libJPEG with a small module that is called
during the DCT and the IDCT computation. We combined our code for
embedding data into the JPEG header for images on static websites and
the modified version of libJPEG into a C++ library that provides
high-level functions for generating and reading of protected images to
our applications.

Our code requires functions to be called from C libraries directly.
Although Gecko engine 2 (the engine underlying Firefox 4) will allow
this directly, the current version of the Gecko engine does not.
As a result, we had to add an XPCOM component to our extension. Using
the XPCOM framework we are able to call routines from native code in
the extension as we would do it with Javascript or XUL functions.
Thus, the final step in our implementation was to extend our library
to be compatible with XPCOM and to add the result as a component to
our Firefox extension.

The image lookup itself is performed by handler functions in the
viewer which check websites for protected images. These functions
check every $\langle img\rangle$-element on a whitelisted website 
(we use a social network whitelist to reduce the overhead introduced
by our technique, see Section~\ref{sec:exp_analysis})
being a protected image, whether it is statically present or
dynamically inserted using Javascript. The latter is the case for most
images on social networks, as their websites are usually created dynamically. If a
protected image is found, the viewer decrypts it and replaces the
$\langle img\rangle$-tag containing the protected image locally with a
new one containing the whole decrypted image data in \emph{base64}
encoded form. Although the use of base64 encoded image data in
$\langle img\rangle$-tags is not part of the HTML standard yet, it is
considered as good as standard and nearly all browsers support this
feature. For the user, this replacement is seamless and does not
require any interaction.

Besides the two client applications, our approach further contains the
keyserver. It is implemented in Scala \cite{Scala}, a programming
language targeting at the Java Virtual Machine (JVM) \cite{JVM}. In
addition, we use the Lift-framework \cite{Lift} for web application
development and PostgreSQL \cite{PSQL}. The latter is used to store
the keys sent by the users. All communication between the
\XPIRE extension and the \KEYSERVERA is implemented using 
XMLHTTPRequests in Javascript. Further, data is transfered using HTTPS, thus
ensuring secure and reliable communication.

To prevent attackers from collecting keys, our implementation permits
the use of \CAPTCHAS as described before. We use Google
reCAPTCHA~\cite{reCAPTCHA} as a concrete implementation. reCAPTCHAs
are based on text extracted from old books which could not be
recognized by OCR tools. The whole communication with the Google
servers is also secured by HTTPS.

\section{Experimental Analysis}
\label{sec:exp_analysis}
Introducing an extra step into the workflow of uploading and viewing
images on the Internet also introduces some extra time needed to
complete the calculations.  We performed several client-side
benchmarks to evaluate the time needed to encrypt and embed images and
to evaluate the overhead that is incurred by the \XPIRE plug-in while
viewing websites. In addition, we compared the two methods of
embedding an encrypted image into a container JPEG: embedding the
payload into the header field and embedding the payload into the
luminance channel.

The benchmarks were run on a notebook with 4GB RAM and a 2.2 GHz Intel
Core 2 Duo CPU using Mac OS X 10.6.4 as the operating system.

To evaluate the time needed for encryption and embedding, we performed
the encryption and embedding computations 50 times for different
numbers of images. The actual time needed for a given number of images
was evaluated as the average over all measurements for a particular
number of images. The results are depicted in Figure
\ref{fig:perf_encrypt}. As expected, embedding data into the header
field is much faster than the luminance channel method. Both methods
scale linearly.

To measure the overhead when viewing websites, we created static
HTML-pages rather than using existing websites. The reason for this is
that a single real world page like Facebook consists of several
elements which are provided by different servers. Moreover, with each
reload, the page looks different and consists of different elements.
Further, if only one server that contributes elements to a website
such as Facebook is not responding, incorrect timing results will be
obtained.

Again, we benchmarked with different numbers of pictures and 50 runs
per series.  Figure \ref{fig:perf_view} and Figure
\ref{fig:perf_decrypt} show the overhead results.  
On a web page which
contains no images that are recognized by the viewer, the overhead of
using the plug-in in comparison to not using the plug-in is not
measurable (Figure~\ref{fig:perf_view}). On websites with
protected images, the luminance channel method is again slower than the
header field method (Figure~\ref{fig:perf_decrypt}).  Fully decrypting
images using the header field method is as fast as checking whether a
possible candidate is really a protected image using the luminance
channel method (Figure~\ref{fig:perf_decrypt}). Neither the header
storing method, nor the bit embedding method scales linearly; 
we expect that this happens due to the way Firefox implements
Javascript DOM handlers and events
\cite{MDN-XUL-EVENTS}.

We also analyzed how the centralized server approach scales in terms
of number of active sessions and number of simultaneous requests. The
machine used as a server was an off-the-shelf office computer with 4GB
RAM and an Intel Core i3 540 CPU with two cores and multithreading
enabled.  We used Ubuntu 10.04 Server Edition as the operating system,
Jetty 6 \cite{Jetty} as the Java webserver, and ApacheBench \cite{AB}
as the benchmark tool.  We assigned a maximum of 2GB heap space to the
JVM running the Jetty instance. Since Jetty does not provide a way to
display the number of active sessions, we can only give rough upper
bounds derived from the total number of session creating requests.
Using this method, we were able to create about 820,000 sessions which
resulted in nearly 2GB of memory usage. At this point, the JVM garbage
collector slowed the system down to the point where the server stopped
responding to our requests. So even when using off-the-shelf equipment,
our \KEYSERVERA is able to handle a reasonable number of users.

In a different benchmark evaluating the server's scalability in terms
of CPU load, we were able to perform about 3000 requests per second
with the same session, effectively reaching a CPU usage between 90\%
and 100\% on all cores according to htop \cite{htop}.

\section{Discussion}
\label{sec:discussion}
\XPIRE has raised a huge discussion in Germany and received an
overwhelming attention in the German media. Reports
about \XPIRE occurred in most of the major printed press, in major TV
channels, and in the radio. The discussion on \XPIRE mainly occurred in
online media.

Most of the criticism of \XPIRE happened because people had wrong 
expectations and knowledge about the functionality of \XPIRE. We provide
our paper now to the public not only to provide in-depth information
to all people interested in this approach, but also to clearly state
once more what \XPIRE can do, and what not. We believe that our approach
is pretty close to what is achievable from a technical point of view.

\vspace{4mm}
\noindent \XPIRE is able to provide the following functionality:
\vspace{2mm}
\begin{compactitem}
\item{Encryption of JPEG images and associating them with an
    expiration date.}
\item{Uploading these images to social networks such as Facebook or Flickr.}
\item{Viewing the images after the upload to a social network with our
    browser plug-in in supported browsers.}
\item{Captchas to increase the cost heavily for one attacker when collecting
    huge amounts of keys.}
\end{compactitem}

\vspace{4mm}
\noindent What \XPIRE is not intended to provide:
\vspace{2mm}
\begin{compactitem}
\item{Does not prevent users from intentionally copying images after
    they got decrypted; this is always possible.}
\item{Users installing malware to collaboratively collect keys and
    store them on a third-party server whenever a picture is viewed;
    this is comparable to intentionally copying images that could also
    be stored unencrypted on another server.}
\end{compactitem}
\vspace{3mm}

Especially the latter is suggested as a break of our system by
\cite{federrath_2011}. However, we never claimed that \XPIRE offers
any form of protection against this kind of attacks. While this kind
of attack could only be hardened, it can never be fully prevented as
it constitutes the technical limit of what can be achieved. Further
protection has to be provided e.g. by law enforcement.

Related to the attack described above, some people have the opinion
that the privacy improvement through \XPIRE is limited. Of course one
would prefer if the aforementioned limitations would be avoided
technically (which is not possible), but, even in the presence of
these limitations, \XPIRE improves the user's privacy heavily. Without
a huge amount of intentionally bad users that collaboratively destroy
the privacy of single users, this system provides a functional
expiration date. This is clearly an improvement in comparison to one
single attacker being able to simply write a crawler to collect all
existing images. Only a huge bunch of either intentionally bad users
or widely installed malware can collect keys or images efficiently.

\section{Conclusion and Future Work}
\label{sec:concl_future}
We have developed a novel, fast, and scalable system called \XPIRE
that allows users to set an expiration date for images in social
networks (e.g., Facebook and Flickr) and on static websites, without
requiring any form of additional interaction with these web pages.
Once the expiration date is reached, the images become unavailable.  A
major technical challenge for rendering the approach possible for
social networks -- where by far the largest number of sensitive images
are published -- was to develop a novel technique for embedding
encrypted information within JPEG files in a way that survives JPEG
compression in real-world implementations. An additional feature of
our approach is that the publishing user can dynamically prolong or
shorten its expiration dates later, and even enforce instantaneous
expiration. We have implemented our system and conducted performance
measurements to demonstrate its efficiency. Our system can be applied
to other data formats such as text messages as well, but we decided to
focus on images first because of their distinguished importance for
user privacy and their underlying technical challenges.  Our current
focus is to support additional social networks and other data formats
in order to increase the usability of the approach.

\bibliographystyle{plainurl}
\bibliography{bib-general}

\appendix
\section{Further information on JPEG}
\label{sec:append_JPEG_Exp}

\begin{figure}[H]
  \begin{center}
    \begin{minipage}[]{0.48\textwidth}
		\includegraphics[width=0.9\textwidth]{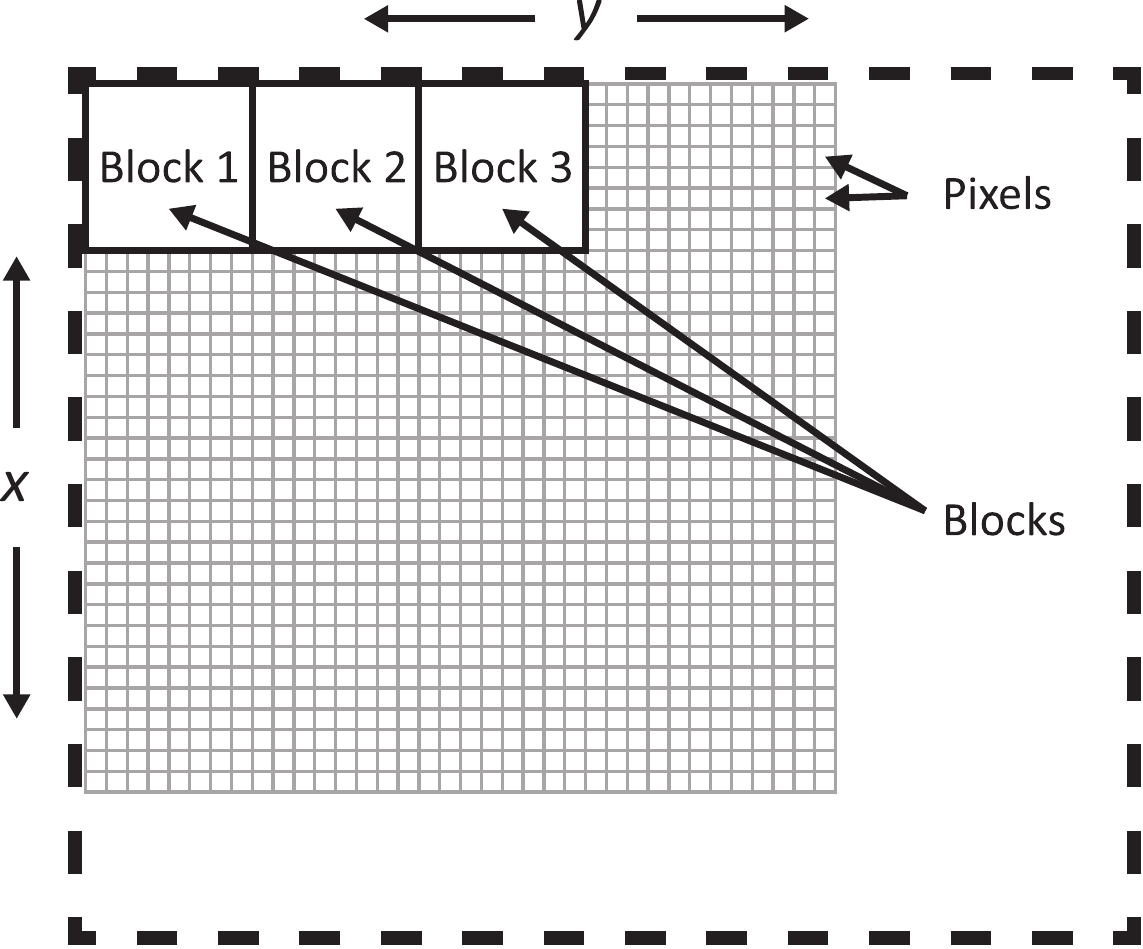}
		\caption{JPEG block extraction}
		\label{fig:jpegblocks}
	\end{minipage}
	\begin{minipage}[]{0.48\textwidth}
		\begin{center}
			  \begin{tabular}[]{ r r r r r r r r }
 
    5 &  3 &  3 &  5 &  7 & 12 & 15 & 18 \\
    4 &  4 &  4 &  6 &  8 & 17 & 18 & 17 \\
    4 &  4 &  5 &  7 & 12 & 17 & 21 & 17 \\
    4 &  5 &  7 &  9 & 15 & 26 & 24 & 19 \\
    5 &  7 & 11 & 17 & 20 & 33 & 31 & 23 \\
    7 & 11 & 17 & 19 & 24 & 31 & 34 & 28 \\
   15 & 19 & 23 & 26 & 31 & 36 & 36 & 30 \\
   22 & 28 & 29 & 29 & 34 & 30 & 31 & 30 \\
   
  \end{tabular}
 
			\caption{Quantization table used by Facebook (luminance channel)}
			\label{fig:fbookqtable}
		\end{center}
	\end{minipage}
  \end{center}
\end{figure}

Figure~\ref{fig:jpegblocks} shows how single blocks are
extracted from a single image.
After the block extraction, a quantization as described in
Section~\ref{sec:jpeg_embedding} is applied. The quantization table
used by Facebook is shown in Figure~\ref{fig:fbookqtable}.

\section{Image quality when using \XPIRE}
\label{sec:appendix:imagequal}
Table~\ref{tab:datasnw} shows the maximum amount of bytes that can be
embedded into the cover images for the aforementioned social networks.
The maximum possible resolution is the maximum pixel size stored by
the social network. All bigger pictures are resized to fit these
limits.  Besides the total amount of image bytes contained in our
cover image, we provide the amount of bytes that can be embedded into
it (including the error correcting codes) and the number of bytes that
can be actually used for storing the image to be embedded. Since
social networks store images only up to a specific resolution, the
maximal file size accepted for embedding is the file size of the image
to be embedded already after it has been resized and re-encoded to a
resolution accepted by upload interfaces of social networks. Hence it
is possible for example to provide a 3500x2700px image that has a file
size of 3.6MB to \XPIRE and after scaling it to 720x720px and
re-encoding it for Facebook it can still be embedded into a cover
image.

\begin{figure*}[]
\begin{minipage}[t]{0.49\textwidth}
    \includegraphics[width=0.93\linewidth]{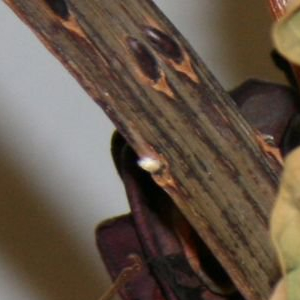}
    \caption{Image uploaded using \XPIRE; it has the identical image quality as the Facebook image on the right side }
    \label{fig:uplourapp}
\end{minipage}
\hfill
 \begin{minipage}[t]{0.49\textwidth}  
    \includegraphics[width=0.93\linewidth]{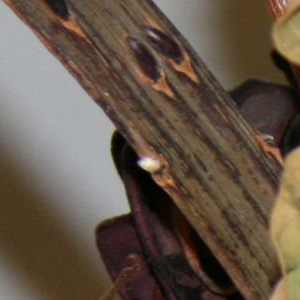}
    \caption{Image uploaded using only Facebook}
    \label{fig:uplfbook}
 \end{minipage}
\end{figure*}

The images below show that the image quality remains unchanged when
our method of uploading an image in a cover image is used.
Figure~\ref{fig:uplourapp} shows a 300x300px area of an image uploaded
using our approach, whereas Figure~\ref{fig:uplfbook} shows the same
area of an image uploaded using only the Facebook interface.

\begin{table}[]
	\begin{centering}
		\begin{tabular} { l | r | r | r | r }
		\toprule
		\textbf{Social Network} & \textbf{max. Resolution} & \textbf{Image Bytes} & \parbox[c][][c]{2cm}{\centering{\textbf{Bytes}}} & \parbox[c][][c]{2cm}{\centering{\textbf{Bytes}}}\\
								&						   &					  & \parbox[c][][c]{2cm}{\centering{incl. ECC}}	     & \parbox[c][][c]{2cm}{\centering{w/o ECC}}       \\
		\midrule
		Facebook & 720x720px & 468000& 117000& 87750\\
		Flickr & 1024x1024px & 976896& 244224& 183168\\
		wer-kennt-wen & 620x620px & 341000& 85250 & 63937.5\\
		\bottomrule
		\end{tabular}
		\caption{Amount of bytes that can be embedded into the cover image}
		\label{tab:datasnw}
	\end{centering}
\end{table}

\end{document}